\documentclass[aps,prd,twocolumn,showpacs,showkeys,groupedaddress]{revtex4}

\usepackage{graphicx}

\usepackage{dcolumn}

\usepackage{bm}
\usepackage{epsfig}
\usepackage{subfigure}

\def\be{\begin{equation}}

\def\ee{\end{equation}}
\def\nuh{\hat{\nu}}

\def\bea{\begin{eqnarray}}
\def\eea{\end{eqnarray}}
\def\nuh{\hat{\nu}}
\def\nue{\nu_e}
\def\num{\nu_{\mu}}
\def\nut{\nu_{\tau}}

\def\gsim{\mathrel{\vcenter{\hbox{$>$}\nointerlineskip\hbox{$\sim$}}}}

\begin{document}

\title{Sterile neutrino signals from supernovae}

\author{
P. Ker\"anen $^{a,b}$,
J. Maalampi$^{a,c}$, 
M. Myyryl\"ainen$^{a}$,
and J. Riittinen$^{a}$}

\affiliation{$^a$ Department of Physics, P.O.~Box~35, FIN-40014 University of Jyv\"askyl\"a, Finland}
\affiliation{$^b$ Radiation and Nuclear Safety Authority, P.O.~Box~14, FIN-00881 Helsinki}
\affiliation{$^c$ Helsinki Institute of Physics, P.O.~Box~64, FIN-00014 University of Helsinki}

\date{\today}

\bigskip\bigskip

\begin{abstract}
We investigate the effects of a mixing of active and sterile neutrinos on the ratios of supernova electron 
neutrino flux ($F_e$) and antineutrino flux ($F_{\bar e}$)  to the total flux of the other neutrino and 
antineutrino flavours ($F_a$). We assume that the heaviest (in the normal hierarchy) Standard Model neutrino 
$\nu_3$ mixes with a sterile neutrino resulting in a pair of mass eigenstates with a small mass gap. Using the density matrix formalism we solve numerically the the evolution of neutrino states in the envelope of a supernova and determine the flux ratios $F_e/F_a$ and $F_{\bar{e}}/F_a$ as a function of the active-sterile mixing angle and for the experimentally allowed range of the standard active-active mixing angle $\theta_{13}$.
\end{abstract}


\maketitle

\noindent{\textit{Introduction.}}
The neutrino oscillation phenomena observed so far can be explained in terms 
of three active neutrinos $\nue, \num$ and $\nut$ (for reviews, see \cite{Strumia}, \cite{Gonzalez-Garcia}). The 
only exception to this has been, until recently,  the muon and electron neutrino and antineutrino oscillation 
claim from LSND experiment \cite{LSND}, whose consistency with the other data would have required the 
existence of a fourth neutrino type, a sterile neutrino. The recent results from MiniBooNE experiment 
\cite{miniboone} seem to refute this claim, however.
Hence, as far as the interpretation of the existing experimental data is concerned there seems to be neither 
need nor too much space for sterile neutrinos \cite{Maltoni}. (Notice, however, Ref. \cite{Goswami}).

Nevertheless, it is quite possible that sterile neutrinos do exist anyway. Many electroweak theories beyond the Standard Model predict the existence of right-handed neutrinos, putting neutrinos in this respect on the same footing as the other basic particles, charged leptons and quarks.  If the right-handed neutrinos exist 
they would be sterile under  the SU(2)$\times$U(1) interactions of the Standard Model (SM). Many Grand 
Unified Theories, e.g. the SO(10) model, include sterile neutrinos, too.  

Obviously, sterile neutrinos would have  escaped detection if their mixing with the active neutrinos is very 
small or non-existing. Another reason for their elusiveness could be that they are closely degenerate in mass 
with the active neutrinos and therefore kinematically indistinguishable from them. The two close-mass 
eigenstates resulting from the mixing of an active neutrino and a sterile neutrino could have appeared in 
experiments performed so far as a single neutrino with the normal SM properties, not as two separate states with 
interactions deviating from the SM. 

We have shown earlier that degenerate pairs of neutrinos resulting in mixing of an active and a sterile 
neutrino, while not separable in laboratory experiments or in the oscillation phenomena studied so far, might 
reveal themselves by affecting the flux rations of the active neutrinos  $\nue, \num$ and $\nut$  measured at 
the Earth and originating in distant astrophysical objects like active galactic nuclei (AGN) \cite{AGN} and 
supernovae \cite{SN}. Neutrino fluxes from these sources are sensitive to smaller squared mass differences 
$\Delta m^2$ than those detected previously.

In this paper we will return to the subject and study in more detail the effects of sterile neutrinos on the flux 
of electron neutrinos ($F_e$) and electron antineutrinos ($F_{\bar{e}}$) from supernovae. We have developed a 
numerical code that computes the evolution of neutrino energy states in a medium of varying density and 
determines the flux ratios $F_e/F_a$ and $F_{\bar{e}}/F_a$, where  $F_a$  is the total flux of muon and tau 
neutrinos and antineutrinos, $F_a=F_{\mu}+F_{\bar{\mu}}+F_{\tau}+F_{\bar{\tau}}$. We find out that even 
quite a small mixing of a sterile and an active neutrino might make the ratios $F_e/F_a$ and in particular 
$F_{\bar{e}}/F_a$ to deviate substantially from  the values they obtain in the case of three active neutrinos.

\vspace{0.5truecm}

\noindent{\textit{The scheme.}}
We will work in the following active-sterile mixing scheme \cite{AGN}. We define 
\be
\nu_{\ell}=U_{\ell i}\nuh_i\quad \quad (i=1,2,3),
\ee
 where $U$ is the standard unitary $3\times 3$ matrix parameterized by three rotation angles $\theta_{12}, 
\theta_{23}$ and $\theta_{13}$ and $\nu_{\ell}$ ($\ell=e,\mu,\tau$) are the ordinary active neutrinos. We 
then assume that each state $\nuh_{i}$ is accompanied by a sterile state $\nu_{si}$, and that these two states mix 
(to be called as an $ii'$-mixing) according to
\bea
\nu_i&=&\cos\phi_i\nuh_i-\sin\phi_i\nu_{si},\nonumber\\
\nu'_i&=&\sin\phi_i\nuh_i+\cos\phi_i\nu_{si},
\label{asmix}\eea
and form two mass eigenstates with masses $m_i$ and $m'_i$, respectively. 
In the limit where the mixing angles $\phi_i$ or the squared mass differences $\Delta 
m^2_{ii'}=m_i^2-m_i^{'2}$, or both, are very small, the rotation angles $\theta_{ij}$ approach the rotation 
angles $\theta_{12}$, $\theta_{23}$ and $\theta_{13}$ of the standard three active neutrino case and 
$\nuh_i$'s  the three ordinary mass eigenstates. 

In this work we will use the following experimental best fit values for the Standard Model  mixing 
angles $\theta_{12}$ and $\theta_{23}$ and the squared mass differences \cite{Gonzalez-Garcia}:
\bea
\theta_{12}&=&33.7^{\circ},\nonumber\\
\theta_{23}&=&43.3^{\circ},\nonumber\\
\Delta m_{12}^2 &=&7.9 \cdot 10^{-5} \rm{eV}^2,\nonumber\\
 \Delta m_{23}^2&=&2.6 \cdot 10^{-3} \rm{eV}^2.
\label{bestfit}\eea
 For the mixing angle $\theta_{13}$ we use the value $\theta_{13}=5.2^{\circ}$, which is the experimental 
upper limit 
 at $1\sigma$ level  ( at $3\sigma$ level $\theta_{13}<11.5^{\circ}$ ) \cite{Gonzalez-Garcia}.

\vspace{0.5truecm}

\noindent{\textit{Solar constraints.}}
Before moving to supernova physics we have to study the restrictions the solar neutrino data sets on the 
active-sterile neutrino mixing we are considering. The SNO solar neutrino experiment \cite{SNO} has 
established  boundaries for the relative neutrino fluxes of the active neutrinos arriving from the Sun 
\cite{Gonzalez-Garcia}. Sterile neutrino mixing would affect these ratios, and by requiring that the effects are 
within the data uncertainties one can constrain the values of  active-sterile mixing parameters
$\phi_i$ and $\Delta m_{ii'}^2$.  Our numerical study shows that solar neutrino data restrict the 
$11'$- and $22'$-mixings strongly leaving room only for mixings  with very
small $\Delta m^2$ or very small mixing angle, which is in good accordance with earlier work done by other 
authors (e.g. \cite{earliersolar}). The restrictions arise mainly from the low energy part of the flux, where the 
vacuum oscillation dominates strongly over the matter effects. 
It turns out that given these tight constraints, the $11'$- and $22'$-mixings would not have noticeable effect on supernova neutrino fluxes.
For the $33'$-mixing, in contrast, there is practically no restrictions from the solar
neutrino data. This is due to the almost negligible  $\nu_e$ proportion in the $\nu_3$ and $\nu'_{3}$ states.

This leaves a possibility for large  $\Delta m_{33'}^2$ and
$\phi_3$, which allows  the active-sterile mixing to have a detectable effect on supernova
neutrino fluxes as we will show in the following. From now on we will assume that sterile states $\nu_{s1}$ 
and $\nu_{s2}$ effectively decouple and that only the sterile state $\nu_{s3}$ affects the behaviour of active 
neutrinos.

\vspace{0.5truecm}

\noindent{\textit{Numerical study.}}
We determine the evolution of the states of neutrinos traversing the supernova envelope by solving numerically the 
density matrix equation
\begin{equation}
\frac{{\rm d}\rho}{{\rm d}t}=i [\rho,H],
\end{equation}
where the Hamiltonian $H$ is given by
\be
H=\frac{1}{2E}U{\rm diag}(  m_1^2,
 m_2^2,
 m_3^2,
m_3^{'2})U^{\dag} + {\rm diag}(  V,0 ,0 ,V/2),
\label{Hamiltonian}
\ee
where $V=\sqrt 2 G_{F}{ N}_e$, and ${ N}_e$ is the number density of electrons \cite{MSW}, assuming that the neutron density is equal to the electron density. The evolution of antineutrino  states is  obtained similarly, except that the matter term $V$ in the Hamiltonian has the opposite sign.

The flavor decomposition of the neutrino flux as a function of time is given by the diagonal elements 
$\rho_{ii}$ of the density matrix once  the initial fluxes are given.

We will use the typical density profile of  a progenitor star as given in \cite{Kachelriess:2001sg}. This yields 
the potential  ($r\simeq t$)
\begin{equation}
V(r)=1.5 \cdot 10^{12}\:  \rm{eV} \: (\rm{m}/r)^3.
\label{eq:kachelpot}
\end{equation}
Following \cite{Fogli:2001pm} we assume that  the progenitor star extends to $R=2.8 \cdot 10^{10}$ m. That 
is, we take $r$ to vary  from the neutrino sphere at about $10^5$ m to $R$.
 
In Fig. 1 the energy states $\nu_i$ ($i=1,2,3$) and $\nu'_3$ are plotted as a function of the electron density inside the star, starting from the surface, where the density of matter and subsequently the potential $V$ become zero. At high 
density the potential $V$ dominates the Hamiltonian (\ref{Hamiltonian}) and the energy eigenstates are close 
to the flavour states $\nu_{\ell}$ and $\bar{\nu_{\ell}}$ ($\ell=e,\mu,\tau,s$). At zero density the states equal to 
the mass eigenstates (\ref{asmix}). 
\begin{figure}[th!]
\centering
\epsfig{file=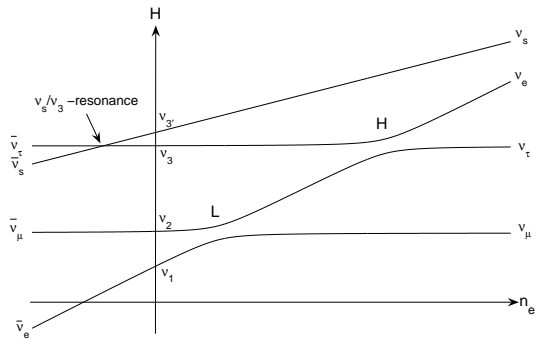,height=5cm}
\caption{Neutrino eigenstates in matter in the presence of 33'-mixing. 
The $L$-resonance is always adiabatic, whereas behavior in the $H$- and 
33'-resonances depend on the value of the corresponding mixing angles 
$\theta_{13}$ and $\varphi$.}
\label{fig:states}
\end{figure}

We assume the initial neutrino and antineutrino
fluxes to be as given in \cite{Raffelt:2003en}. The authors of that reference have developed a model for 
supernova explosion  and based on this model they give the fluxes of $\nu_e$, 
$\bar{\nu}_{e}$ and $\nu_{\mu}\simeq\nu_{\tau}\simeq\bar\nu_{{\mu}}\simeq\bar\nu_{{\tau}}\equiv\nu_x$ as a 
function of neutrino energy. More specifically, we will use in our analysis the ratios 
$F_e^0:F_{\bar{e}}^0:F_x^0=4:2.3:1.4$, which correspond to the fluxes at the energy $10$ MeV where the 
intensity of the neutrino radiation is largest. Of course, the fluxes, $F_x^0$ in particular, will be hard to measure within a single energy bin \footnote{The energy spectum of $\nu_x$ might be possible to measure in the future experiments, see e.g. \cite{clean}}. Nevertheless, the flux ratios at the peak energy $10$ MeV turn out to be very close to the integrated fluxes, with just slightly overweighting the fraction of electron neutrinos, so that they can be reliably used as input fluxes. Sterile neutrinos are of course not produced in supernovae but they are created, as a result of their assumed mixing with the standard model neutrinos, during the traverse of neutrinos the medium of the star.

\vspace{0.5truecm}

\noindent{\textit{Results.}}
Let us first assume that sterile neutrinos are absent and study the system with the three active neutrinos  and 
their antineutrinos only. 
By using these as initial conditions we solved numerically the density matrix equation and determined the the 
flux ratios $F_e/F_a$ and $F_{\bar{e}}/F_a$. Allowing the mixing angle $\theta_{13}$ to vary in the range 
$0\leq\theta_{13}\leq 5,2^{\circ}$ and fixing the other mixing parameters to their best-fit values (\ref{bestfit}) the flux ratios in the case of three active neutrinos are found to vary within the ranges 
\bea
0.17<&F_e/F_a&<0.29,\nonumber\\
0.24<&F_{\bar{e}}/F_a&<0.26.
\label{exratios}
\eea

The sensitivity of these ratios on the mixing angle $\theta_{13}$ is due to the fact that the adiabaticity of the 
so called $H$-resonance  at the crossing of the energy states $\nu_2$ and $\nu_3$ (see Fig. \ref{fig:states}) 
depends on the value of this angle. When $\theta_{13}=0$, the behavior of the system  is fully non-adiabatic, 
meaning that the flux of electron neutrinos, which is the largest one initially, ends to the energy state $\nu_2$, 
which contains less of $\nu_a$ (= $\nu_{\mu}+ \nu_\tau+\bar{\nu}_{\mu}+\bar{\nu}_{\tau}$) than the energy state $\nu_3$. 
This corresponds to the largest values of the ratios $F_e/F_a$ and $F_{\bar{e}}/F_a$. The smallest values of these 
ratios are obtained when $\theta_{13}$ attains its largest experimentally allowed value. In this case the 
$H$-resonance is adiabatic and the initial electron neutrino flux ends into the state $\nu_3$, which has a very 
small $\nu_{e}$ component, $\vert\langle\nu_e\vert\nu_3\rangle\vert^2=\sin^2\theta_{13}=0.008$. 

We now add into the scheme a sterile neutrino. We assume that the sterile neutrino mixes with the active 
neutrinos in such a way that the states $\nu_1$ and $\nu_2$ remain intact, that is, their flavor decomposition 
is the same as in the Standard Model and consist only of active flavors, while the third state $\nu_3$ becomes 
a superposition of the active neutrinos and the sterile neutrino according to (\ref{asmix}). The state $\nu_3$ is 
assumed to be close in mass to the new heavier state $\nu'_3$ orthogonal to it. In our analysis we will 
assume $m_3^{'2}-m_3^{2}=10^{-6}$ eV$^2$. Increasing $\Delta m_{33'}^2$ from this value would not change our results. For lower values of $\Delta m_{33'}^2$, the effect of active-sterile mixing decreases and will be washed out at 
$\Delta m_{33'}^2 \approx 10^{-11} \rm{eV}^2$, where our results would coincide with the SM result. As $\nu'_3$ is heavier than $\nu_3$, the energy levels corresponding to these two states cross in the antineutrino sector, not in the neutrino sector (see Fig. \ref{fig:states}).

\begin{figure}[h!]
\centering
\subfigure[$F_{\bar{e}}/F_a$]
{\epsfig{file=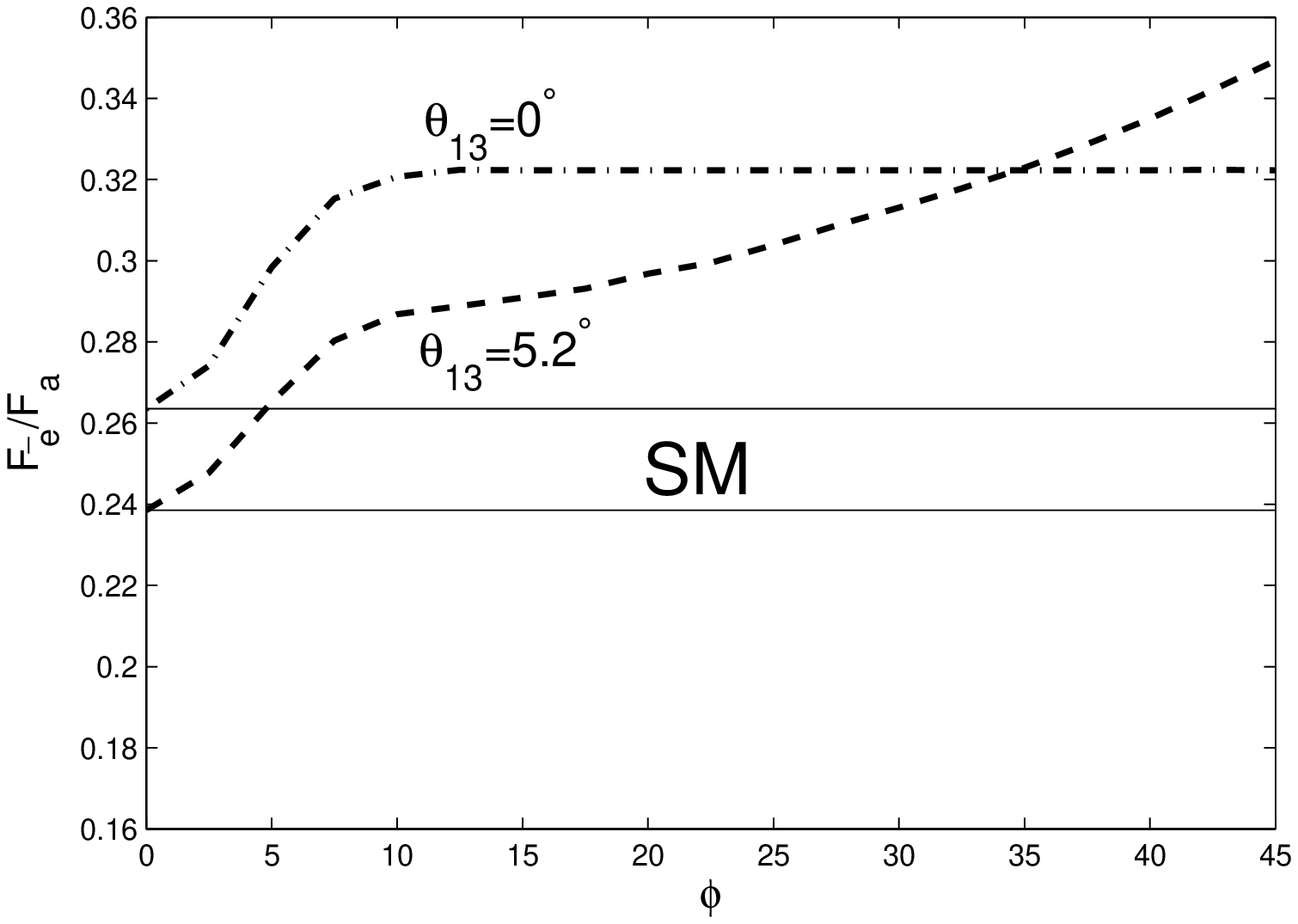,width=7cm}\label{fig:sub:a}}
\vspace{0.15cm}
\subfigure[$F_e/F_a$] 
{\epsfig{file=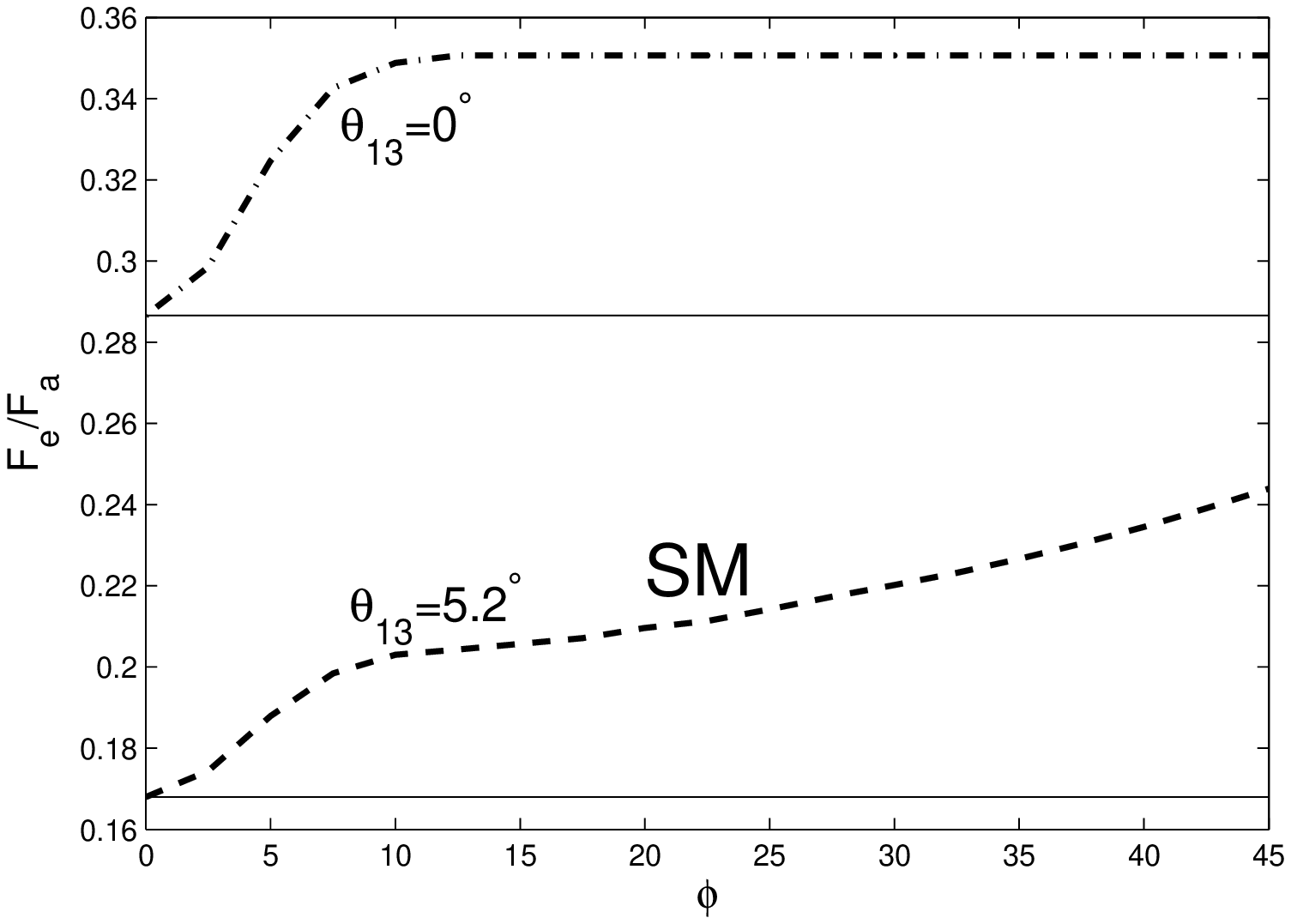,width=7cm}\label{fig:sub:b}}
\caption{Ratio of \subref{fig:sub:a} $\bar{\nu}_e$
\subref{fig:sub:b} $\nu_e$ and nonelectronic active fluxes as 
a function of sterile mixing angle $\varphi$. Results corresponding to the
minimum and maximum values of $\theta_{13}$ allowed by experiment are shown. The region denoted by SM 
is the allowed range
in the SM case of three active neutrinos.
 Here the neutrino energy is taken as $E=10$ MeV.}
\label{fig:vacosc} 
\end{figure}

In Fig. \ref{fig:vacosc} we present the flux ratios $F_e/F_a$ and $F_{\bar{e}}/F_a$ as a function of the 
active-sterile mixing angle $\phi$ for  the experimentally allowed range of the angle  $\theta_{13}$, 
$0\leq\theta_{13}\leq 5,2^{\circ}$. In what follows we will explain the behavior of the ratios qualitatively.

The both ratios are larger for $\theta_{13}=0^{\circ}$ than for $\theta_{13}=5.2^{\circ}$. This feature is easy 
to understand in the following way. When 
$\theta_{13}=0$, the $H$-resonance in the neutrino sector is non-adiabatic and none of the electron
neutrinos created in the supernova core ends up to the state $\nu_3$ and the state  $\nu_3$ will be 
populated solely by the less intense $\nu_\tau$ component of the initial flux.  The original electron neutrinos 
end up to the state $\nu_2$ whose $\nu_e$ component is much larger than that of $\nu_3$, where the 
original electron neutrinos will end to in the case of  $\theta_{13}=5.2$. As to the ratio  $F_{\bar{e}}/F_a$, the 
flux $F_{\bar{e}}$  does not depend on the value of $\theta_{13}$, but  $F_a$ is larger  in the case of 
$\theta_{13}=5.2^{\circ}$ than in the case  of  $\theta_{13}=0^{\circ}$ as then the original electron neutrino 
flux ends to the state $\nu_3$, where there are more muon and tau neutrinos than in the state $\nu_2$.

The 33'-mixing affects the ratios $F_e/F_a$ and $F_{\bar{e}}/F_a$ in two ways. Firstly, the mixing makes the 
flavor decomposition of the energy eigenstate $\nu_3$ to change from its SM form as it replaces  the active 
flavors partly by the sterile flavor. As there is only a tiny  electron neutrino (antineutrino) component in 
$\nu_3$ ($\bar\nu_3$), this would affect mainly the flux $F_a$ and makes the value of the ratios $F_e/F_a$ 
and $F_{\bar{e}}/F_a$ larger than  in the Standard Model case. The effect of course increases with increasing  
active-sterile mixing angle $\phi$. This explains the growth of the ratios when $\phi $ shifts from zero to about 
$12^{\circ}$ . 

The second effect of the active-sterile mixings is related to the possible level crossing in the resonance 
transition region of the $\bar\nu_3$ and $\bar\nu'_3$ antineutrino states. For a small mixing angle $\phi$, the 
33'-transition is non-adiabatic and the level crossing takes place. That is, 
all $\bar\nu_{\tau}$'s created in the supernova will end up to the state $\bar\nu_3$, which has only a tiny 
sterile content, proportional to $\sin^2 \phi$.  

When $\phi>12^{\circ}$  the 33'-resonance is fully adiabatic. In this case the original $\nu_{\tau}$ flux will go 
into the state $\nu'_3$, which has a sterile fraction proportional to $\cos^2 \phi$. When $\phi $ increases, 
the $\bar\nu_{\mu}$ and $\bar\nu_{\tau}$ components of  $\bar\nu'_3$  will become larger, which will make 
$F_a$ to {\it increase}. What will happen to the ratios depends, however, on the adiabaticity of the 
H-resonance. When $\theta_{13}=0$ the H-resonance is non-adiabatic, the occupation of the state $\nu_3$ 
originates from the original $\nu_{\tau}$ flux.  When the angle $\phi$
increases, the $\nu_{\mu}$ and $\nu_{\tau}$ components of  $\nu_3$ will become smaller and the sterile part will become larger. This will make  $F_a$ to  {\it decrease}. As the original flux of tau neutrinos and antineutrinos 
are assumed to be equal, the net effect will be that  $F_a$ remains practically constant for $\phi>12^{\circ}$. 
As the electron neutrino and antineutrino fluxes are quite insensitive to the active-sterile mixing, this implies 
that the values of the ratios $F_e/F_a$ and $F_{\bar{e}}/F_a$ saturate to the constant values for 
$\phi>12^{\circ}$ and $\theta_{13}=0$. According to  Fig. \ref{fig:vacosc}, the saturation values are 
$F_e/F_a=0.35$ and $F_{\bar{e}}/F_a= 0.32$, to be compared with  the Standard Model values 0.29 and 
0.26, respectively. 

When $\theta_{13}=5.2^{\circ}$,  the H-resonance is adiabatic. Neutrinos created originally as $\nu_e$ will in 
this case end 
up to the state $\nu_3$. In contrast with the previous case, now the ratios $F_e/F_a$ and $F_{\bar{e}}/F_a$ 
will not saturate after the active-sterile mixing has achieved the value $\phi=12^{\circ}$ but will keep  
increasing with $\phi$. The difference to the previous case is that now the state $\nu_3$ will be more 
occupied than the state $\nu'_3$ as the original electron neutrino flux is assumed to be larger than the tau 
antineutrino flux. For the maximal 33'-mixing, i.e. for $\varphi=45^{\circ}$, one has  $F_e/F_a= 0.24$ and 
$F_{\bar{e}}/F_a=0.35$, to be compared with the SM values 0.17 and 0.24, respectively. Note that 
$F_{\bar{e}}/F_a$ becomes insensitive to the angle $\theta_{13}$ when the active-sterile mixing angle has the 
value about $35^{\circ}$. 

\vspace{0.5truecm}

\noindent{\textit{Discussion and summary.}
As one can infer from the Fig. \ref{fig:vacosc}, the ratio $F_{\bar{e}}/F_a$ would give more information on the 
active-sterile mixing than $F_e/F_a$ taking the uncertainty of the value of the mixing angle $\theta_{13}$ into 
account. Given the value of $F_{\bar{e}}/F_a$ from measurements, the angle $\phi$ will be limited in general 
into  quite a narrow range of values. Also, for $\phi\gsim 5^{\circ}$ the $F_{\bar{e}}/F_a$ would deviate from 
the SM prediction for all allowed values of $\theta_{13}$, in contrast with $F_e/F_a$.

As for the uncertainties in the initial neutrino fluxes, the differences between the SM case and sterile case arise only due to the $33'$-mixing in the antineutrino part. So only the initial flux of $\bar{\nu}_{\tau}$ will affect the ratios by possibly changing the value of $F_a$.
 
It is worthwhile to make a remark concerning the effect of the Earth on the fluxes. 
As there is only a tiny  electron neutrino component in the state $\nu_3$, the effects of neutrino-matter interactions in the Earth will not differ much from those in the SM case. The active-sterile mixing will only slightly change the fluxes of non-electron flavour neutrinos  measured at the opposite side of the Earth. The most visible effect would be a distortion of the $\nu_l$, ($l \neq e,\bar{e}$) energy spectra, which could possibly be measured in
future supernova neutrino experiments \cite{clean}. 

In summary, we have considered the situation where neutrinos appear as three pairs of states, where each 
pair orginates in a mixing of an active neutrino and a sterile neutrino and its components are so close in mass 
that they are not distinguished as different particles in the present experiments. We have shown that this kind 
of structure might manifest itself through the fractions of electron neutrinos and antineutrinos  in the 
supernova neutrino flux,  i.e. through the ratios $F_e/F_a$ and $F_{\bar{e}}/F_a$.  Apart from the 
active-sterile mixing angle $\phi$, the deviations of these ratios from their Standard Model values are sensitive 
also to the mixing angle $\theta_{13}$ for which an upper limit only is known at the moment. Only an 
active-sterile mixing associated with the heaviest mass state $\nu_3$ can have measurable effects, since the 
solar neutrino data restrict the mixing parameters in the case of the two other states too small to lead any 
observable changes in the supernovae neutrino fluxes.

\begin{acknowledgments}

This work has been supported by the
Academy of Finland under the contracts no.\  104915 and 107293. One of us (MM) expresses her gratitute to 
the Finnish Cultural Foundation for a grant.
\end{acknowledgments}

\end{document}